\documentclass[aps,preprint,prb,showpacs]{revtex4-1}
\usepackage{graphicx,amssymb,amsmath}
\usepackage{graphicx}
\usepackage{dcolumn}
\usepackage{bm}

\newcommand{\nt}{$\nu_T=1$}
\newcommand{\dl}{$d/\ell$}
\newcommand{\bpar}{$B_{||}$}
\newcommand{\bperp}{$B_{\perp}$}
\newcommand{\Sxx}{$\sigma_{xx}^{||}$}

\begin{document}

\title{Exciton Condensation and Perfect Coulomb Drag}

\author{D. Nandi$^1$, A.D.K. Finck$^1$, J.P. Eisenstein$^1$, L.N. Pfeiffer$^2$, and K.W. West$^2$}

\affiliation{$^1$Condensed Matter Physics, California Institute of Technology, Pasadena, CA 91125
\\
$^2$Department of Electrical Engineering, Princeton University, Princeton, NJ 08544}

\date{\today}


\maketitle

{\bf
Coulomb drag is a process whereby the repulsive interactions between electrons in spatially separated conductors enable a current flowing in one of the conductors to induce a voltage drop in the other \cite{pogrebinski,price,rojo}.  If the second conductor is part of a closed circuit, a net current will flow in that circuit. The drag current is typically much smaller than the drive current owing to the heavy screening of the Coulomb interaction.  There are, however, rare situations in which strong electronic correlations exist between the two conductors.  For example, bilayer two-dimensional electron systems can support an exciton condensate consisting of electrons in one layer tightly bound to holes in the other \cite{fertig89,jpemacd,macdgirvin}.  One thus expects ``perfect'' drag; a transport current of electrons driven through one layer is accompanied by an equal one of holes in the other \cite{su}. (The electrical currents are therefore opposite in sign.)  Here we demonstrate just this effect, taking care to ensure that the electron-hole pairs dominate the transport and that tunneling of charge between the layers is negligible.}

The exciton condensate of interest here develops at high perpendicular magnetic field $B_{\perp}$ when the separation $d$ between two parallel two-dimensional electron systems (2DESs) is comparable to the magnetic length $\ell = (\hbar/eB_{\perp})^{1/2}$ and the total density $n_T=n_1+n_2$ of electrons in the bilayer (we consider only the balanced case $n_1=n_2$) matches the degeneracy $eB_{\perp}/h$ of a single spin-resolved Landau level \cite{fertig89,jpemacd,macdgirvin}.  Hence, the total Landau level filling factor is $\nu_T = 1$. When $d/\ell \lesssim 1.8$ an energy gap to charged excitations opens and the bilayer electron system displays a quantized Hall (QH) plateau $\rho_{xy}=h/e^2$.  The interlayer tunneling conductance becomes strongly enhanced near zero bias and, when equal electrical currents are driven in opposite directions through the two layers, the Hall effect vanishes at low temperature \cite{spielman1,kellogg2,tutuc1,wiersma}.  

Even in the limit of zero tunneling through the barrier separating the layers, interlayer Coulomb interactions at \nt\ are sufficient to both open the charge gap and to spontaneously generate quantum phase coherence between electrons in opposite layers.  Spontaneous interlayer phase coherence allows the system ground state to be described as a Bose condensate of interlayer excitons.  At low temperatures the charged excitations are frozen out and unable to transport current across the bulk of the 2D system.  In contrast, the neutral electron-hole pairs in the condensate remain populous and free to move about the bulk.  Transport of these excitons is equivalent to counterflowing electrical currents in the two layers. 

Evidence for exciton transport at \nt\ was obtained from Hall effect measurements in which counterflowing electrical currents were driven through the two layers \cite{kellogg2,tutuc1,wiersma}.  The observed vanishing of the Hall voltage at \nt\ is consistent with the counterflowing currents being carried by excitons.  However, interpretation of these experiments is complicated by the conducting edge states which exist at the boundary of all quantum Hall systems.  Furthermore, since the experiments were performed using simply-connected Hall bar geometries, they were incapable of proving that excitons were moving through the bulk of the 2DES.  Subsequent experiments by Tiemann {\it et al} and later Finck {\it et al}, employed the multiply-connected Corbino geometry (essentially an annulus with separated edge states on the two rims) in order to search for exciton transport across the insulating bulk \cite{tiemann1,tiemann2,finck}.  By connecting the two layers together at one rim while applying a voltage between the layers at the other rim, Finck {\it et al.} observed that relatively large, oppositely directed currents would flow across the bulk of the \nt\ QH state \cite{finck}. This observation contrasts sharply with the observed inability of the bulk to support co-directed currents in the two layers.  Finck {\it et al} concluded that bulk exciton transport was responsible for their results \cite{finck}.   

We show here that the excitonic correlations built into the \nt\ QH state can force oppositely directed currents to flow in the two layers even when there is no electrical connection between them.  For small driving currents the observed drag current is closely equal in magnitude to the drive current; i.e. the drag is ``perfect''.  The role of interlayer tunneling is investigated and shown to be irrelevant in the proper circumstances.  

Our sample consists of two parallel 2DESs confined in a GaAs/AlGaAs double quantum well structure, the details of which are given below.  The bilayer 2DES is patterned into an annulus (1 mm inner, 1.4 mm outer diameter) with arms extending from each rim to ohmic contacts; a schematic plan-view of the device is shown in Fig. 1a.  Each ohmic contact may be connected either to both 2D layers simultaneously or to either layer separately \cite{sepcon}.  Electrostatic gating of the 2DESs in the annulus (but not the contact arms) allows the key parameter $d/\ell$ to be tuned from $d/\ell= 2.35$ down to about  $d/\ell =1.49$ at \nt.  In addition to the perpendicular magnetic field \bperp\ needed to establish \nt, an in-plane field \bpar\ may also be applied by tilting the sample relative to the total magnetic field.  This allows us to suppress \cite{spielman2} the interlayer tunneling which can otherwise pollute Coulomb drag measurements.  The sample displays a robust QH effect at \nt\ for $d/\ell \lesssim 1.8$ for all tilt angles up to at least $\theta = 66^\circ$.  Figure 1a illustrates the insulating character of the bulk of the 2D system at \nt\ (with $d/\ell=1.5$ and $\theta =26^\circ$) in an Arrhenius plot of the Corbino conductance $\sigma_{xx}^{||}$ for parallel transport in the two layers.  A small ac excitation voltage ($V_{ex}\sim 18$ $\mu$V at 13 Hz) is applied between contacts (to both layers) on the inner and outer rim of the annulus.  The resulting current flow $\delta I$, plus the rim-to-rim voltage difference $\delta V$ between two additional contacts, are recorded and used to compute $\sigma_{xx}^{||}=\partial I/\partial V $.  As expected, the conductance is thermally activated, $\sigma_{xx}^{||} \sim e^{-\Delta/2T}$, and gives an energy gap $\Delta \approx 360$ mK.  When a dc bias $V_{dc}$ is added to the ac excitation voltage $V_{ex}$, the conductance \Sxx\ increases.  This smooth ``breakdown'' of the \nt\ QH effect, shown in Fig. 1b, has important consequences for the Coulomb drag results to which we now turn.

In a Corbino Coulomb drag measurement, a voltage $V$ is again applied between the inner and outer rims of the annulus, but only via contacts to one of the two 2D layers. The other layer is either left open or is closed upon itself by connecting an external resistor between the two rims.  The open-circuit case is similar to the \Sxx\ measurement discussed above; current will flow in the drive layer, but only in proportion to \Sxx. The closed-circuit case is potentially different; if strong interlayer correlations are present, relatively large {\it oppositely directed} currents, mediated by exciton transport, might flow in the two layers.  

Figure 2 shows the results of such closed-circuit drag measurements.  External resistors in both the drive and drag loops allow us to monitor the currents $I_1$ and $I_2$ flowing in each.  This arrangement is illustrated in the inset to Fig. 2a; $I_1$ and $I_2$ are defined as positive if they flow in the direction of the arrows.  While the drive circuit is grounded at one point, the drag circuit is left to float.  Figure 2a shows the dc currents $I_1$ and $I_2$ flowing, at $T\approx 17$ mK, in response to a dc drive voltage $V_{dc}$ at \nt, with $d/\ell=1.49$ and $\theta = 26^\circ$.  For small $V_{dc}$, $I_1$ and $I_2$ are very nearly equal and grow steadily, if somewhat super-linearly, with voltage.  In the $V_{dc} \rightarrow 0$ limit, the conductances $\partial I_1/\partial V \approx \partial I_2/\partial V$ exceed the parallel flow Corbino conductance \Sxx\ at \nt\ by a factor of 5.  At large $V_{dc}$, the currents separate, with $I_1$ continuing to grow steadily while $I_2$ begins to saturate.  Figure 2b shows the magnetic field dependence (at $\theta = 26^\circ$) of the drag transconductance $\partial I_2/\partial V$ at $V_{dc}=0$ (obtained by applying a weak purely ac excitation voltage $V_{ex}$ across the drive circuit) at $T \approx 25$ mK.  As expected, significant drag is found only in the vicinity of \nt\ at $B_\perp = 1.87$ T.    

Since the currents $I_1$ and $I_2$ are detected outside the bilayer 2DES, it is not obvious that the drive current is truly passing through the top 2D layer and the drag current through the bottom 2D layer.   As noted previously \cite{tiemann1,tiemann2}, interlayer tunneling could be shunting the current from the top to the bottom layer near the outer rim of the device, through external drag layer loop, and then back from the bottom to top layer near the inner rim.  To investigate this possibility, we have examined the tunneling conductance in our sample at both $\theta = 0$ and $26^\circ$.

Figure 3 shows dc tunneling current-voltage characteristics at \nt\ and $d/\ell=1.49$, for $\theta = 0$ and $26^\circ$.  These data were obtained at $T\approx 20$ mK by applying an external dc voltage $V_{dc}$ between contacts to the ``upper'' and ``lower'' 2DES layer on the outside rim of the annulus and recording both the resultant tunneling current $I$ and the interlayer voltage $V_{int}$ between the two remaining outside rim contacts.  Figure 3a shows the tunneling currents plotted versus $V_{int}$ while Fig. 3b plots the currents versus $V_{dc}$; the two figures therefore contrast the ``four-terminal'' and ``two-terminal'' tunneling current-voltage characteristics of the bilayer system.  The four-terminal $I$-$V_{int}$ characteristic at $\theta = 0$ clearly shows the Josephson-like near-discontinuity at $V_{int} = 0$ reported previously \cite{spielman1,spielman2,tiemann2,tiemann3}.  In contrast, the two-terminal $I$-$V_{dc}$ characteristic shows the tunneling current initially rising smoothly with $V_{dc}$.  This difference is due almost entirely to the extrinsic series resistances presented by the arms leading into the annulus.  Indeed, comparison of the two- and four-terminal tunneling data allows us to accurately estimate the series resistances and their non-linearity with voltage; these estimates are important in the analysis of the Coulomb drag data.  Note that for $\theta =0^\circ$ the maximum tunneling currents are comparable to the currents $I_1$ and $I_2$ observed in the drag measurement and shown in Fig. 2a.
  
Most importantly, Figs. 3a and 3b reveal the expected \cite{spielman2} suppression of the tunneling current resulting from tilting the sample.  At $\theta = 26^\circ$ the zero bias anomaly so prominent at $\theta = 0$ in the four-terminal $I$-$V_{int}$ characteristic is essentially obliterated. Even at the relatively high applied voltage of $|V_{dc}| = 300$ $\mu$V the tunnel current is only $\sim 0.1$ nA.  The two- and four-terminal characteristics at $\theta = 26^\circ$ are very similar since the tunneling resistance at this tilt angle is much larger than the extrinsic series resistances.

Comparing the tunneling data in Fig. 3 with the Coulomb drag data in Fig. 2a demonstrates that tunneling is not an important contributor to the drag current $I_2$ at $\theta = 26^\circ$.   Ignoring, for the moment, the different excitation means in the two cases (interlayer vs. intralayer biasing), it is clear that the tunneling conductance near $V_{dc}=0$ is at least 10 times smaller than the drag transconductance $\partial I_2/\partial V$.  Similarly, the drag current at $V_{dc} = 200$ $\mu$V is $I_2\approx 650$ pA, while the tunneling current at the same bias is only 44 pA. This comparison, however, greatly exaggerates the importance of tunneling precisely because of the different excitation means in the two experiments.  In a drag measurement, the bias voltage $V_{dc}$ is applied across the drive layer; the drag layer is allowed to float.  Direct measurements of the interlayer voltage under these conditions shows it to be quite small ($<15$ $\mu$V), even when the intralayer drive voltage reaches $|V_{dc}| \sim 300$ $\mu$V.  At such small interlayer voltages the tunneling current (at $\theta = 26^\circ$) is extremely small ($\lesssim 4$ pA).  Note that this argument fails at $\theta =0$ where, as Fig. 3a shows, a large ($> 1$ nA) and virtually discontinuous jump in the tunneling current occurs at zero interlayer voltage.  

The above discussion enables us to conclude that the drive and drag currents shown in Fig. 2b do indeed flow across the bulk of the top and bottom 2D layers in the annulus, and in opposite directions.  For small drive voltages, $I_1\approx I_2$ and thus the drag is essentially perfect.   In this nearly pure counterflow regime, the drag process is dominated by neutral exciton transport.     

Figures 4a and 4b shows how the drag ratio $I_2/I_1$ depends on temperature $T$ and effective layer separation \dl, respectively.   For small drive voltages $V_{dc}$ the drag ratio is close to unity ($I_2/I_1 \approx 0.97$) only at the lowest $T$ and \dl, where the \nt\ QH state is strongest.  Increasing either parameter reduces the drag ratio at small $V_{dc}$. In all cases, the drag ratio also falls with increasing drive voltage.  We believe that these deviations from perfect drag are due primarily to the finite Corbino conductance \Sxx\ which allows parallel charge transport across the bulk to occur alongside the neutral exciton transport.

Assuming \Sxx=0 and dissipationless exciton transport, Su and MacDonald \cite{su} predict $I_1=I_2=V/(R_1+R_2)$, where $R_1$ and $R_2$ are the net resistances in series with the Corbino annulus in the drive and drag circuits, respectively.  These include the external circuit resistors $R_{ext}$, the resistances $R_{arm}$ of the 2DES arms leading into the annulus, and quantum Hall ``contact'' resistances $R_c$ of order $h/e^2$.  Generalizing the Su-MacDonald model to include non-zero \Sxx, we find, assuming linear response, the drag current reduced to $I_2=V/(R_1+R_2+R_1R_2\sigma_{xx}^{||})$ and the drag ratio to $I_2/I_1=1/(1+R_2\sigma_{xx}^{||})$.  Using the tunneling data in Fig. 3a to estimate $R_1$ and $R_2$ and the measured \Sxx\ data in Fig. 1a, we can estimate the expected drag ratio $I_2/I_1$ near zero bias.  These estimates, shown in Fig. 4a, compare quite favorably with the observed drag ratios at $T = 17$, 35, and 50 mK.  At higher bias the non-linearity of \Sxx\ (shown in Fig. 1b) and the series resistances $R_1$ and $R_2$ must also be taken into account.  The dashed lines in Fig. 2b display the results of one such calculation of the drag and drive currents. The qualitative agreement with the experimental results is strong evidence that the enhanced Corbino conductance \Sxx\ at elevated temperatures and drive voltages is the dominant source of deviations from perfect Coulomb drag at \nt.

The assumption that exciton transport across the bulk of the 2DES is dissipationless can be questioned in real, disordered samples such as ours \cite{huse,fertig,fil1,cooper1}.  A phenomenological excitonic ``resistance'' $R_s$ can be introduced whereby $R_s(I_1+I_2)$ equals the difference $\Delta V_{int}$ between the interlayer voltages on the two rims of the annulus (a spatially uniform interlayer voltage would produce no dissipation).  In the \Sxx=0 limit, this new resistance leaves the drag perfect, but reduces the currents to $I_1=I_2=V/(R_1+R_2+R_s)$.  The relatively large magnitude of $R_1+R_2$ (never less than $2h/e^2$ \cite{su,pesin}) limits the ability of the present Coulomb drag experiments to detect small values of $R_s$.  Future multi-terminal measurements should be able to set stringent limits on any dissipation occurring in the exciton channel.

{\bf Methods}  The present sample consists of two 18 nm GaAs quantum wells separated by a 10 nm Ga$_{0.1}$Al$_{0.9}$As barrier.  The center-to-center layer separation is therefore $d=28$ nm. This double well structure is flanked by thick Ga$_{0.7}$Al$_{0.3}$As cladding layers.  Si doping sheets within the cladding layers populate the lowest subband of each quantum well with a 2DES of nominal density $5.5 \times 10^{10}$ cm$^{-2}$ and low temperature mobility of $1 \times 10^6$ cm$^2$/Vs. Standard photo-lithographic techniques are used to pattern the bilayer 2DES into the geometry depicted in Fig. 1.  Diffused AuNiGe ohmic contacts are positioned at the ends of arms extending away from both rims of the annulus.  Electrostatic gates cross these arms (both on top and thinned backside of the sample) in order to employ a selective depletion scheme which allows the contacts to communicate with the 2DES in the annulus either via both layers in parallel or either layer separately \cite{sepcon}.  Additional gates control the 2D layer densities in the annulus itself.  The sample is mounted on a Ag platform in good thermal contact with the mixing chamber of a $^3$He-$^4$He dilution refrigerator.   The electrical transport measurements reported here employ standard dc and/or low frequency ac techniques.  The dc drive and drag currents $I_1$ and $I_2$ can be determined either by exciting the drive circuit with a purely dc voltage, or by numerical integration of the ac currents $\delta I_1$ and $\delta I_2$ flowing in response to a weak ac voltage ($V_{ex} \sim 18$ $\mu$V at 13 Hz) added to the dc component of $V$.  While the two methods are in excellent agreement, the ac technique is less noisy.

{\bf Acknowledgements -} 
We thank A.H. MacDonald and D. Pesin for discussions.  This work was supported via NSF grant DMR-1003080.

{\bf Author contributions -} 
DN, ADKF, and JPE conceived the project.  LNP and KWW grew the samples. DN and ADKF performed the experiment and, along with JPE, analyzed the data and wrote the manuscript.

{\bf Additional information -} 
The authors declare no competing financial interests.

\begin{figure}[h]
\begin{center}
\includegraphics[width=4.5in] {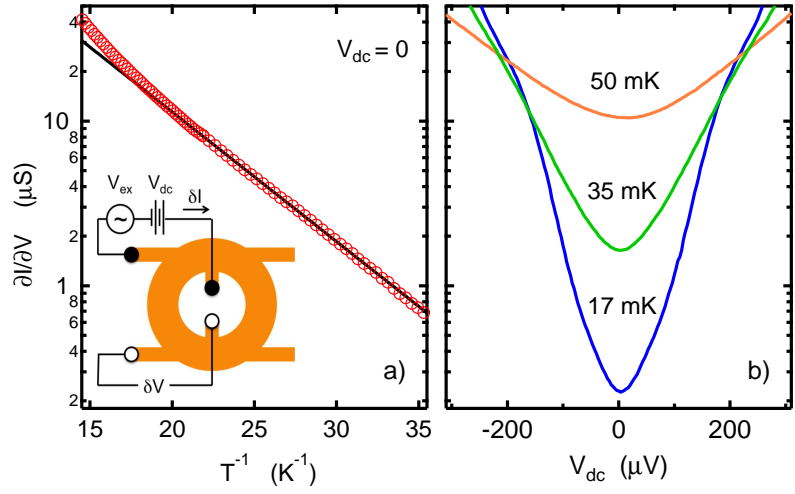}
\end{center}
\caption{a) Corbino conductance \Sxx\ = $\partial I/\partial V$ at zero dc bias vs. $T^{-1}$ at \nt\ with \dl\ = 1.5 and $\theta = 26^\circ$.  Solid line implies an energy gap of $\Delta \approx 360$ mK.  Inset: Schematic plan-view of device.  Solid dots indicate contacts to both layers; open dots contacts to lower layer only.  b) Corbino conductance $\partial I/\partial V$ vs. applied dc bias $V_{dc}$ under same conditions as in a) for various temperatures.}
\label{fig1}
\end{figure}

\begin{figure}
\begin{center}
\includegraphics[width=4.5in] {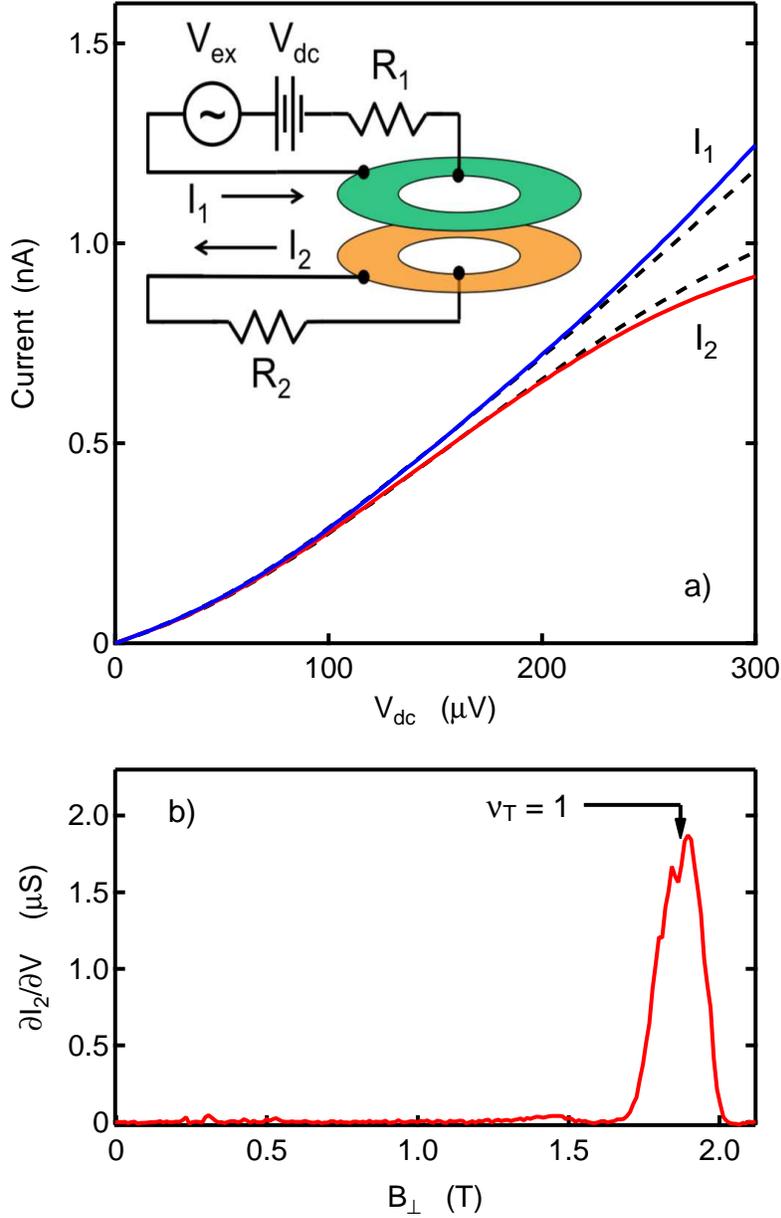}
\end{center}
\caption{Corbino Coulomb drag.  a) Drive and drag currents, $I_1$ and $I_2$, respectively, vs. dc bias $V_{dc}$ at \nt, with \dl\ = 1.5 and $T = 17$ mK.   Dashed lines are simulations incorporating estimated series resistances and measured Corbino conductvity \Sxx.  Inset depicts measurement circuit.  The resistors $R_1$ and $R_2$ comprise both external circuit resistors and resistances intrinsic to the device.
b) Drag transconductance, $\partial I_2/ \partial V$ at $V_{dc}=0$ vs. perpendicular magnetic field at $T\approx 25$ mK.  All data at $\theta = 26^\circ$.}
\label{fig2}
\end{figure}

\begin{figure}
\begin{center}
\includegraphics[width=4.5in] {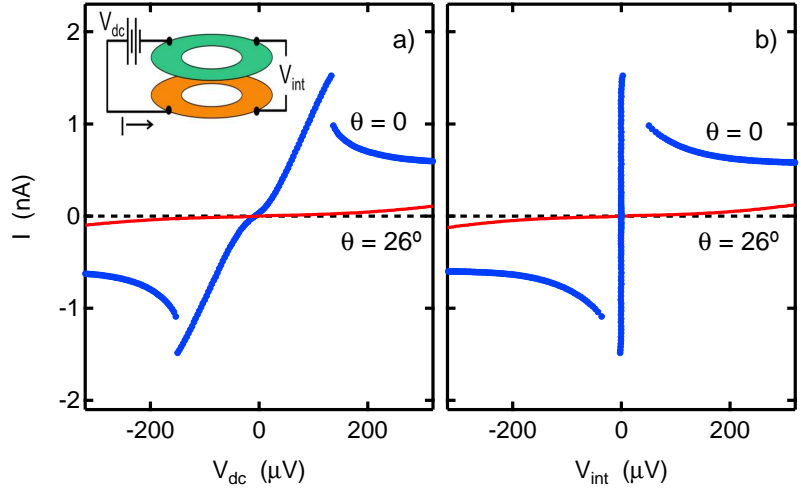}
\end{center}
\caption{Four-terminal vs. two-terminal tunneling characteristics at \nt\ for $\theta=0$ and $26^\circ$.  a) Tunnel current vs. measured interlayer voltage $V_{int}$.  b) Same as a) except tunnel current is plotted vs. applied interlayer bias $V_{dc}$.  All data at \dl\ = 1.49 and $T \approx 20$ mK. }
\label{fig3}
\end{figure}

\begin{figure}
\begin{center}
\includegraphics[width=4.5in] {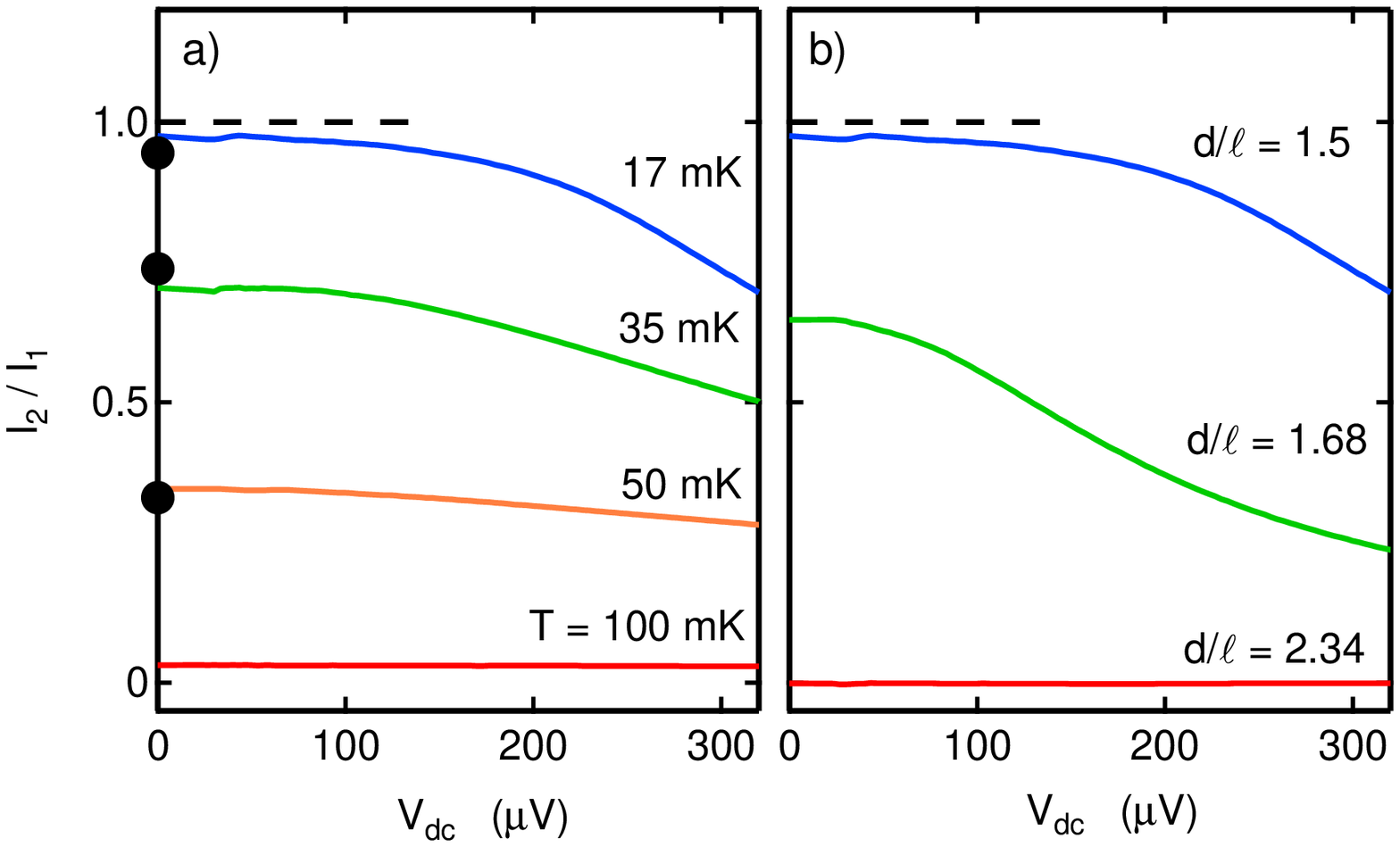}
\end{center}
\caption{a) Drag ratio, $I_2/I_1$ vs. dc bias $V_{dc}$ at \nt\ for $d/\ell=1.5$ at various temperatures.  Solid dots are estimates based on measurements of \Sxx\ at $V_{dc}=0$, as described in text. b) Drag ratio vs. $V_{dc}$ at \nt\ and $T=17$ mK for various \dl.  All data at $\theta = 26^\circ$.}
\label{fig4}
\end{figure}


\begin{references}

\bibitem{pogrebinski} Pogrebinsky, M.B. Mutual drag of carriers in a semiconductor-insulator-semiconductor system. {\it Fiz. Tekh. Poluprovodn.} {\bf 11}, 637-644 (1977) [{\it Sov. Phys. Semicond.} {\bf 11}, 372-376 (1977)].

\bibitem{price} Price, P.M. Hot electron effects in heterolayers. {\it Physica (Amsterdam)} {\bf117B}, 750-752 (1983).

\bibitem{rojo} Rojo, A.G. Electron-drag effects in coupled electron systems. {\it J. Phys. Condens. Matter} {\bf 11}, R31-R52 (1999).

\bibitem{fertig89} Fertig, H.A. Energy spectrum of a layered system in a strong magnetic field. {\it Phys. Rev. B} {\bf40}, 1087-1095 (1989).

\bibitem{jpemacd} Eisenstein, J.P \& MacDonald, A.H. Bose-Einstein condensation of excitons in bilayer electron systems. {\it Nature} {\bf 432}, 691-694 (2004) and the references cited therein.

\bibitem{macdgirvin} MacDonald, A.H. \& Girvin, S.M. in {\it Perspectives in Quantum Hall Effects}, edited by S. Das Sarma and A. Pinczuk (Wiley, New York, 1997).

\bibitem{su} Su, Jung-Jung \& MacDonald, A.H. How to make a bilayer exciton condensate flow. {\it Nature Physics} {\bf 4}, 799-802 (2008).

\bibitem{spielman1} Spielman, I.B., Eisenstein, J.P., Pfeiffer, L.N. \& West, K.W.  Resonantly Enhanced Tunneling in a Double Layer Quantum Hall Ferromagnet. Phys. Rev. Lett. {\bf 84}, 5808-5811 (2000).

\bibitem{kellogg2} Kellogg, M., Eisenstein, J.P., Pfeiffer, L.N. \& West, K.W. Vanishing Hall Resistance at High Magnetic Field in a Double-Layer Two-Dimensional Electron System. {\it Phys. Rev. Lett.} {\bf 93}, 036801 (2004).

\bibitem{tutuc1} Tutuc, E., Shayegan, M. \& Huse, D.A. Counterflow Measurements in Strongly Correlated GaAs Hole Bilayers: Evidence for Electron-Hole Pairing. {\it Phys. Rev. Lett.} {\bf 93}, 036802 (2004).

\bibitem{wiersma} Wiersma, R. {\it et al.} Activated Transport in the Separate Layers that Form the $\nu_T=1$ Exciton Condensate. {\it Phys. Rev. Lett.} {\bf 93}, 266805 (2004).

\bibitem{tiemann1} Tiemann, L. {\it et al.} Exciton condensate at a total filling factor of one in Corbino two-dimensional electron bilayers. {\it Phys. Rev. B} {\bf 77}, 033306 (2008).

\bibitem{tiemann2} Tiemann, L., Dietsche, W., Hauser, M. \& von Klitzing, K. Critical tunneling currents in the regime of bilayer excitons.  {\it New J. of Phys.} {\bf 10}, 045018 (2008).

\bibitem{finck} Finck, A.D.K., Eisenstein, J.P., Pfeiffer, L.N. \& West, K.W. Exciton Transport and Andreev Reflection in a Bilayer Quantum Hall System. {\it Phys. Rev. Lett.} {\bf 106}, 236807 (2011).

\bibitem{sepcon} Eisenstein, J.P., Pfeiffer, L.N. \& West, K.W. Independently contacted two-dimensional electron systems in double quantum wells. {\it Appl. Phys. Lett.} {\bf 57}, 2324-2326 (1990). 

\bibitem{spielman2} Spielman, I.B., Eisenstein, J.P., Pfeiffer, L.N. \& West, K.W.  Observation of a Linearly Dispersing Goldstone Mode in a Quantum Hall Ferromagnet. {\it Phys. Rev. Lett.} {\bf 87}, 036803 (2001).

\bibitem{tiemann3} Tiemann, L., Yoon, Y., Dietsche, von Klitzing, K. \& Wegscheider, W. Dominant parameters for the critical tunneling current in bilayer exciton condensates. {\it Phys. Rev. B} {\bf 80}, 165120 (2009).

\bibitem{huse} Huse, David A. Resistance due to vortex motion in the $\nu =1$ bilayer quantum Hall superfluid. {\it Phys. Rev. B} {\bf 72}, 064514 (2004).

\bibitem{fertig} Fertig, H.A. \& Murthy, Ganpathy. Coherence Network in the Quantum Hall Bilayer. {\it Phys. Rev. Lett.} {\bf 95}, 156802 (2005).

\bibitem{fil1} Fil, D.V. \& Shevchenko, S.I. Interlayer tunneling and the problem of superfluidity in bilayer quantum Hall systems. {\it Low Temp. Phys.} {\bf 33}, 780-782 (2007).

\bibitem{cooper1} Lee, D.K.K., Eastham, P.R. \& Cooper, N.R. Breakdown of counterflow superfluidity in a disordered quantum Hall bilayer. {\it Adv. Condens. Matter Phys.} {\bf 2011}, 792125 (2011).

\bibitem{pesin} Pesin, D.A. \& MacDonald, A.H. Scattering theory of transport in coherent quantum Hall bilayers. {\it Phys. Rev. B} {\bf 84}, 075308 (2011).

\end{references}
\end{document}